\documentclass[
    ,final            
  ]
  {aipproc}

\layoutstyle{8x11double}

\begin{document}

\title{Non-thermal radiation from molecular clouds illuminated by cosmic rays from nearby supernova remnants.}


\author{Stefano Gabici}{
  address={Dublin Institute for Advanced Studies, 31 Fitzwilliam Place, Dublin 2, Ireland}
}

\author{Sabrina Casanova}{
  address={Max--Planck--Institute f\"ur Kernphysik, Saupfercheckweg 1, 69117 Heidelberg, Germany}
}

\author{Felix A. Aharonian}{
  address={Dublin Institute for Advanced Studies, 31 Fitzwilliam Place, Dublin 2, Ireland}
  ,altaddress={Max--Planck--Institute f\"ur Kernphysik, Saupfercheckweg 1, 69117 Heidelberg, Germany} 
}

\begin{abstract}
Molecular clouds are expected to emit non-thermal radiation due to cosmic ray interactions in the dense magnetized gas. Such emission is amplified if a cloud is located close to an accelerator of cosmic rays and if cosmic rays can leave the accelerator and diffusively reach the cloud. We consider the situation in which a molecular cloud is located in the proximity of a supernova remnant which is accelerating cosmic rays and gradually releasing them into the interstellar medium. We calculate the multiwavelength spectrum from radio to $\gamma$-rays which emerges from the cloud as the result of cosmic ray interactions. The total energy output is dominated by the $\gamma$-ray emission, which can exceed the emission from other bands by an order of magnitude or more. This suggests that some of the unidentified TeV sources detected so far, with no obvious or very weak counterpart in other wavelengths, might be associated with clouds illuminated by cosmic rays coming from a nearby source. 
\end{abstract}

\maketitle



Galactic Cosmic Rays (CRs) are believed to be accelerated via first order Fermi mechanism at the expanding shocks of supernova remnants (SNRs). 
In this context, a detection of SNRs in TeV $\gamma$-rays was predicted by \cite{dav1}, due to the decay of neutral pions produced in interactions between the accelerated CRs and the interstellar gas swept up by the SNR shock.

The detection of several SNRs in $\gamma$-rays \cite{jim}, though encouraging, cannot provide by itself the final proof that CRs are indeed accelerated at SNR shocks. This is because competing leptonic processes, namely inverse Compton scattering from accelerated electrons, can also explain the observed TeV emission, provided that the magnetic field does not significantly exceed $\sim 10 ~ \mu$G. Evidence for strong $\approx 100 \mu$G magnetic fields, and thus indirect support to the hadronic scenario for the $\gamma$-ray emission, comes from the observation of thin X-ray synchrotron filaments surrounding some SNRs \citep{bamba,vink,heinz} and of the rapid variability time scale of the synchrotron X-rays~\cite{iasunobu}.
A conclusive proof of the hadronic nature of the $\gamma$-ray emission will possibly come from the detection of neutrinos, which are produced during the same hadronic interactions responsible for the production of $\gamma$-rays \citep{dav1}.

The presence of a massive molecular cloud (MC) close to a SNR can provide a dense target for CR hadronic interactions and thus enhance the expected $\gamma$-ray emission. 
If the MC is overtaken by the SNR shock, the $\gamma$-ray emission is expected to be cospatial with the SNR shell, or a portion of it. If the MC is located at a given distance $d_{cl}$ from the SNR, it can still be illuminated by CRs that escape from the SNR and emit $\gamma$-rays~\cite{atoyan,gabici2}. For this scenario, it has been shown 
that, for typical SNR parameters and for a distance $D = 1$~kpc, a MC of mass $10^4 M_{\odot}$ emits TeV $\gamma$-rays at a detectable level if it is located within few hundreds pc from the SNR~\cite{gabici2}. In this case the angular displacement between the SNR and the $\gamma$-ray emission is $\approx 6^{\circ} (D/1 {\rm kpc})^{-1} (d_{cl}/100 {\rm pc})$. 
This translates in the fact that sometimes the association between SNRs and MCs can be not so obvious, given  that the separation between the two objects can be bigger than the instrument field of view. Following this rationale, it was proposed 
\cite{gabici2} that some of the unidentified TeV sources \cite{hessunid,milagrounid} 
might be MCs illuminated by nearby SNRs.

In this paper, we calculate the expected non-thermal emission, from radio to multi-TeV photons, from a MC illuminated by CRs coming from a nearby SNR. We generalize the model presented in \cite{gabici2}, which was limited to the hadronic TeV photons only, to include the generation of secondary electrons in the cloud and the related synchrotron and Bremsstrahlung emission components.
Electrons accelerated at the SNR shock remain confined in the shell due to severe synchrotron losses.
Similar calculations have been recently published, though they are focused on the $\gamma$-ray \cite{diego} and radio \cite{protheroe} emission only.
We found that the total radiation energy output from a MC is dominated by the $\gamma$-ray emission, which can exceed the emission from other bands by an order of magnitude or more. This reinforces the belief that some of the unidentified TeV sources detected so far, with no obvious or very weak counterparts in other wavelengths (the so called "dark sources"), might be in fact associated with massive MCs illuminated by CRs. Moreover, under certain conditions, the $\gamma$-ray spectrum from the cloud exhibit a concave shape, being steep at low ($\sim$ GeV) energies and hard at high ($\sim$ TeV) energies. This fact might have important implications for the studies of the spectral compatibility of GeV and TeV $\gamma$-ray sources.



\subsection{Cosmic ray spectrum at the cloud location}

\begin{figure}  
\resizebox{0.4\textwidth}{!}{
\includegraphics{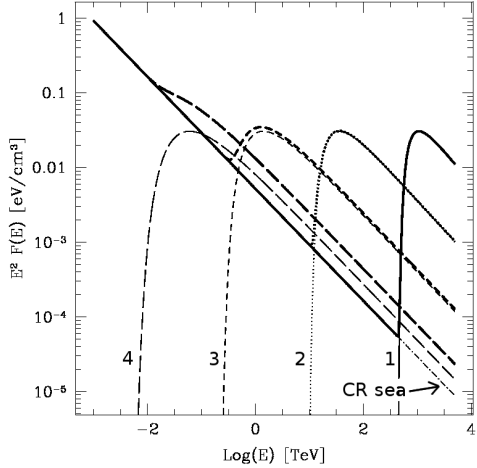}} 
\caption{CR spectrum in a MC located at 100 pc from a SNR. The thin dot--dashed line shows the Galactic CR spectrum. The thin solid (1), dotted (2), short--dashed (3) and long-dashed (4) lines represent the spectrum of CRs coming from the SNR for 500, 2000, 8000 and 3200 yr after the supernova explosion. Thick lines show the total CR spectrum.}
\label{fig:crs}
\end{figure}

We consider a MC located at a given distance from a SNR and we calculate the CR spectrum at the cloud location. 
The spectrum is the sum of two components: {\it i)} the CRs coming from the SNR, and {\it ii)} the galactic CR background, assumed to be constant throughout the Galaxy and equal to the one measured close to the Sun.
While the latter contribution is constant in time, the former changes, since the flux of CRs escaping from the SNR evolves in time (full details are given in \cite{gabici2,gabici3}).
Fig. \ref{fig:crs} shows the CR spectrum at the location of the MC. The Galactic CR background is plotted as a thin dot--dashed line (labeled as CR sea), while the spectrum of the CRs coming from the SNR is plotted as thin lines for different times after the supernova explosion: 500 (solid, 1), 2000 (dotted, 2), 8000 (short--dashed, 3) and 32000 yr (long--dashed, 4).
Thick lines represent the sum of the two contributions.  
The distance between the SNR and the MC is 100 pc.
The total supernova explosion energy is $10^{51}$ ergs and the CR acceleration efficiency is $30 \%$. 
The diffusion coefficient is
$
D(E) = 10^{28} (E/10 ~ GeV)^{0.5} ~ cm^2/s ~ ,
$
compatible with CR propagation models \cite{CRbook}.

The evolution with time of the CR spectrum at the position of the MC can be understood by recalling that CRs with different energies leave the SNR at different times \cite{ptuskin}. 
The highest energy ($\sim$ PeV) CRs leave the SNR first, while CRs with lower and lower energy are released at later times. 
Moreover, higher energy CRs diffuse faster, thus the spectrum of CRs at the MC exhibit a sharp low energy cutoff at an energy $E_{low}$, which moves to lower and lower energies as time passes. 
The position of the cutoff represents the energy of the least energetic particles that had enough time to reach the MC. 

Here we assume that CR can freely penetrate the cloud. For a detailed discussion of this issue see \cite{gabici1}.



\subsection{Non-thermal radiation from the cloud}

The non-thermal emission from a MC located in the proximity of a SNR is the result of CR interactions with the dense gas and magnetic field in the MC and is made up of two contributions: a steady state contribution from the interactions of background CRs that penetrate the cloud and a time dependent contribution from the interactions of CRs coming from the nearby SNR.

\begin{figure}  
\resizebox{.79\textwidth}{!}{
\includegraphics{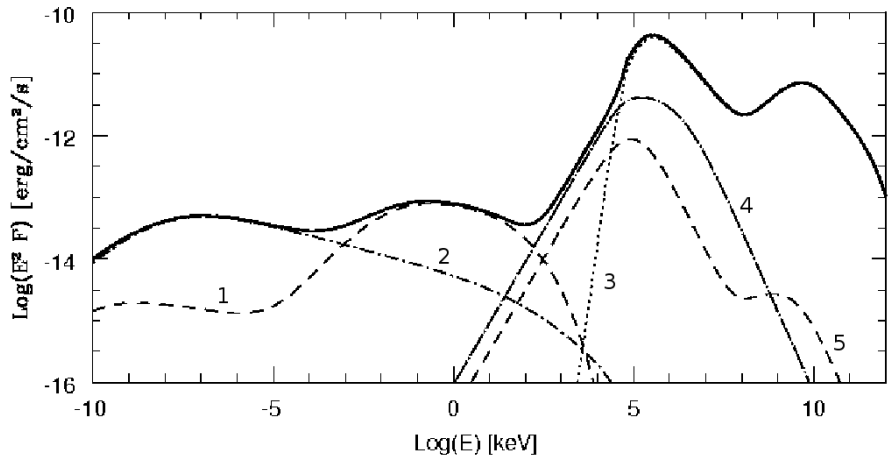}
} 
\caption{Broad band spectrum for a MC of mass $10^5 M_{\odot}$, radius $20$~pc, density $\sim 120$~cm$^{-3}$, magnetic field $20 \mu$G. The MC is at 100 pc from a SNR that exploded 2000 yr ago. Distance is 1 kpc. The dotted line shows the emission from $\pi^0$--decay (curve 3), the dot-dashed lines represent the synchrotron (curve 2) and Bremsstrahlung (curve 4) emission from background CR electrons that penetrate the MC and the dashed lines the synchrotron (curve 1) and Bremsstrahlung (curve 5) emission from secondary electrons.}
\label{fig:GMC2000}
\end{figure}

We consider a MC of mass $10^5 M_{\odot}$, radius $20$~pc and uniform density $\sim 120$~cm$^{-3}$. The magnetic field is $20 \mu$G. 
In order to show all the different contributions to the total non-thermal emission, in Fig.~\ref{fig:GMC2000} we plot the broad band spectrum from the MC for $2000$ yr after the supernova explosion.
The SNR is 100~pc away from the MC. The distance to the observer is 1~kpc.
The dotted line (curve 3) represents the emission from neutral pion decay (from both background CRs and CRs from the SNR), the dot--dashed lines represent the synchrotron (curve 2) and Bremsstrahlung (curve 4) emission from background CR electrons that penetrate the MC and the dashed lines represent the synchrotron (curve 1) and Bremsstrahlung (curve 5) emission from secondary electrons produced during inelastic CR interactions in the dense gas \cite{gabici3}.

The decay of neutral pions dominates the total emission for energies above $\approx 100$~MeV. The two peaks in the emission reflects the shape of the underlying CR spectrum, which, as illustrated in Fig.~\ref{fig:crs}, is the sum of the steep background CR spectrum, which produces the $\pi^0$--bump at a photon energy of $m_{\pi^0}/2 \sim 70$~MeV (in the photon flux $F$), and an hard CR component coming from the SNR that produces the bump at higher energies. The flux level at 1 TeV is approximatively $5 \times 10^{-12}$erg/cm$^2$/s, well detectable by currently operating Imaging Atmospheric Cherenkov Telescopes, even taking into account the quite extended ($\approx 2^{\circ}$) nature of the source. It is remarkable that such a MC would be detectable even if it were located at the distance of the Galactic centre, as can be easily estimated by taking into account that the sensitivity of a Cherenkov telescope like H.E.S.S. after 50 hours of exposure, is $\approx 10^{-13} (\theta_s/0.1^{\circ})$TeV/cm$^2$/s, where $\theta_s$ is the source extension. This means that very massive clouds can be used to reveal the presence of enhancements of the CR density in different locations throughout the whole Galaxy. Similar conclusions can be drawn for the expected GeV emission, which is currently probed by the AGILE and GLAST satellites. In particular, GLAST, with a point source integral sensitivity of $\approx 10^{-9}$GeV/cm$^{2}$/s at energies above 1 GeV (www-glast.slac.stanford.edu), is expected to detect such MCs as extended sources if they are within $1 \div 2$ kpc from the Earth, or as point sources if they are at larger distances.

The spectral shape in the $\gamma$-ray energy range deserves further discussion. For the situation considered in Fig.~\ref{fig:GMC2000}, the $\approx 1 \div 100$ GeV $\gamma$-rays are the result of the decay of neutral pions produced by background CRs that penetrate the MC. Thus, the $\gamma$-ray spectrum at those energies simply mimic the underlying CR spectrum, which is a steep power law  of the form $\approx E^{-2.75}$.
On the other hand, the neutral pion decay spectrum at energies above $\approx 100$ GeV is, in this case, dominated by the contribution from CRs coming from the nearby SNR. After 2000 yr from the supernova explosion, only CRs with energies above several tens of TeV had enough time to leave the SNR and reach the MC and thus the CR spectrum at the cloud exhibits an abrupt low energy cutoff at that energy, that we call here $E^{cut}_{CR}$ (see Fig.~\ref{fig:crs}). As a consequence, the $\gamma$-ray spectrum is expected to be peaked at an energy $\sim E_{CR}^{cut}/10$ of several TeV. The slope of the $\gamma$-ray spectrum below the peak is determined by the physics of the interaction only, and not by the shape of the underlying CR spectrum, and is of the form $\propto E^{-1}$ \cite{kelner}. 
Thus, a loose association between a SNR and a MC is expected to be characterized, at least at some stage of the SNR evolution, by a very peculiar spectrum which is steep at low (GeV) energies and hard at high (TeV) energies.

The possibility of detecting sources with such a peculiar spectrum is also relevant for the issue of identifying GeV and TeV unidentified sources.  
One of the criteria generally adopted to support an association between a GeV and TeV source is, beside the positional coincidence, the spectral compatibility \citep[see e.g.][]{funk}.
Such a criterium would not be applicable to the case studied here.

\begin{figure}  
\resizebox{.79\textwidth}{!}{
\includegraphics{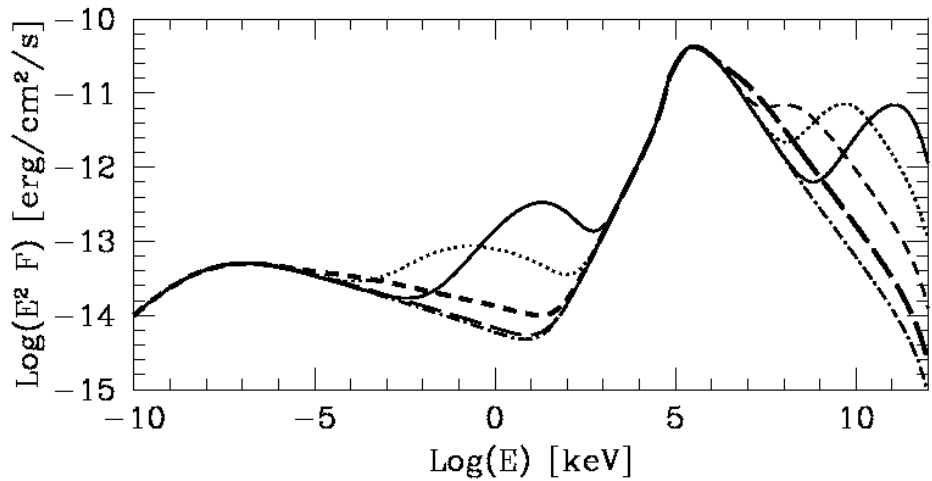}
} 
\caption{Radiation from the MC considered in Fig.~\ref{fig:GMC2000}. The solid, dotted, short-dashed and long-dashed line shows the spectrum at 500, 2000, 8000 and 32000 yr after the explosion. The dot-dashed line is the spectrum of a MC with no SNR in its proximity.}
\label{fig:GMCtot}
\end{figure}

The evolution with time of the emission from the MC is shown in Fig.~\ref{fig:GMCtot}, where the solid, dotted, short--dashed and long--dashed lines show the spectrum for 500, 2000, 8000 and 32000 yr after the supernova explosion respectively. 
For comparison, the emission from a MC with no SNR located in its proximity is plotted as a dot--dashed line. In this case only background CRs that penetrate the MC contribute to the emission.

Fig.~\ref{fig:GMCtot} shows that the radio ($\lambda >  0.1$mm) and the soft $\gamma$-ray ($\approx 0.1 \div 1$ GeV) emissions are constant in time. This is because such emissions are produced by background CRs that penetrate the MC. The emission in the other energy bands is variable in time, being produced by the CRs coming from the SNR whose flux changes with time (see Fig.~\ref{fig:crs}). The most prominent features in the variable emission are two peaks, in X- and $\gamma$-rays respectively. 

The $\gamma$-ray peak moves to lower and lower energies with time, reflecting the fact that CRs with lower and lower energies reach the MC as time progresses.
At early times, the emission can extend up to $\sim$ 100 TeV, revealing the presence of PeV CRs and thus indirectly the fact that the nearby SNR is acting as a CR pevatron \cite{gabici2}.
Moreover, the $\gamma$-ray emission in the TeV range is enhanced with respect to the one expected from an isolated MC (dot--dashed line) for a period of several $10^4$ yr. This is much longer than the period during which SNRs are effectively accelerating the multi-TeV CRs responsible for the TeV emission, which lasts a few thousands years. This is because the duration of the $\gamma$-ray emission from the MC is determined by the time of propagation of CRs from the SNR to the MC and not by the CR confinement time in the SNR. Therefore, the $\gamma$-ray emission from the MC lasts much longer than the emission from the SNR, making a detection more probable~\citep{gabici2}.

The peak in the X-ray spectrum is due to synchrotron emission from secondary electrons produced in CR interactions in the MC. The peak is moving to lower energies with time but, unlike the $\gamma$-ray peak, it is also becoming less and less pronounced. This can be explained as follows: 500 yr after the supernova explosion (Fig.~\ref{fig:GMCtot}, solid line), PeV CRs from the SNR reach the MC and produce there secondary electrons with energy in the $\approx 100$ TeV range. For these electrons, the synchrotron cooling time is comparable with the escape time from the MC \cite{gabici3}. Thus, they release a considerable fraction of their energy in the form of X-ray synchrotron photons before leaving the MC. As time passes, lower energies CRs reach the MC and secondary electrons with lower energies are produced. For these electrons the cooling time becomes progressively longer than the escape time \cite{gabici3} and this explains the suppression of the synchrotron emission.
The X-ray synchrotron emission is weaker than the TeV emission for any time and for times $> 2000$yr the ratio between TeV and keV emission can reach extreme values of a few tens or more.
These values are observed from some of the unidentified TeV sources detected by H.E.S.S. \citep{suzaku1, suzaku2}
In the scenario presented in this paper, spectra showing a high TeV/kev flux ratio can be produced very naturally if a MC is illuminated by CRs coming from a nearby SNR.
This suggestion is supported by the fact that most of the unidentified TeV sources are spatially extended, as MCs are expected to be.

\textit{Acknowledgments:}
SG \& SC acknowledge support from the European Community under a Marie Curie Intra-European-Fellowship.



\bibliographystyle{aipprocl} 



\end{document}